\documentclass{aa}
\usepackage{graphics}
\usepackage{epsfig}
\usepackage{natbib}
\begin{document}
\title{Temperature Isotropization in Solar Flare Plasmas due to 
the Electron Firehose Instability}
\titlerunning{Temperature Isotropization due to Electron Firehose Instability}
\author{Peter Messmer}
\authorrunning{P. Messmer}
\institute{Institute of Astronomy, ETH Zentrum, 8092 Z\"urich, Switzerland}
\offprints{Peter Messmer,\\ \email{messmer@astro.phys.ethz.ch}}
\date{Received 7 August 2001 / Accepted 7 November 2001}

\abstract{
The isotropization process of a collisionless plasma with an
electron temperature anisotropy along an external magnetic field 
($T_\| ^e\gg T_\perp^e$, $\|$ and $\perp$ with respect to the
background magnetic field) and isotropic protons is investigated using a
particle-in-cell~(PIC) code. Restricting wave growth mainly parallel to
the external magnetic field, the isotropization mechanism is
identified to be the Electron Firehose Instability (EFI). The free
energy in the electrons is first transformed into left-hand circularly 
polarized transverse low-frequency waves by a non-resonant interaction. 
Fast electrons can then be scattered towards higher perpendicular 
velocities by gyroresonance, leading finally to a complete
isotropization of the velocity distribution. During this phase of the
instability, Langmuir waves are generated which may lead to the
emission of radio waves. A large fraction of the  protons is resonant with the
left-hand polarized electromagnetic waves, creating a 
proton temperature anisotropy  $T_\|^p < T_\perp^p$. 
The parameters of the simulated plasma are chosen compatible to 
solar flare conditions. The results indicate the significance of this 
mechanism in the particle acceleration context: The EFI limits the anisotropy 
of the electron velocity distribution, and thus provides the necessary
condition for further acceleration. It enhances the pitch-angle of the
electrons and heats the ions.
}
\maketitle

\section{Introduction}
The energization of large numbers of particles to high energies is 
still an unsolved problem in solar flare research. The vast diversity of 
observational data imposes severe constraints, for which any acceleration
model has to account. The current models can broadly be assigned to three
different categories:
{\em Shock acceleration}, {\em acceleration by parallel electrical fields}
and {\em stochastic acceleration by cascading MHD waves}. For a review
of these mechanisms, see e.g. \citet{miller97}. 

One  restriction on the acceleration model is imposed by the time
scales at which the process has to operate: 
Hard X-ray observations revealed spiky structures as short as 
$\approx 400$ ms \citep{kiplinger84, machado93}. Assuming a
thick-target emission process for the radiation, the electrons have
to be accelerated to energies above $20$ keV within the given time in 
each of these spikes. It is believed that these \lq energy release
fragments\rq~are the basic constituent of the entire impulsive phase
\citep{machado93}. Further fragmentation of the energy 
release and electron acceleration process  was proposed based on 
decimetric radio emission \citep{benz85}, with temporal
structures well below $\approx 10$ ms.


Assuming an isotropic electron velocity distribution,  
\citet{miller96} showed that stochastic acceleration can account for
energization of electrons from thermal energies up and above 
$20$keV in less than $400$ ms. However, the actual particle 
acceleration mechanism in this model,  Transit-Time Damping
\citep{fisk76, stix92}, increases only the velocity component parallel to the 
external magnetic field. In absence of an additional pitch-angle
scattering mechanism, this would lead to an anisotropic electron velocity
distribution, reducing the efficiency of the acceleration mechanism.
Due to the low density and the high temperature, the mean free path 
between Coulomb collisions in a solar flare plasma is large. The 
scattering mechanism has therefore to be of a different nature, 
presumably a wave-particle interaction.
 
Predominant acceleration along an external magnetic field $B_0$
compared to the acceleration in perpendicular direction is not only a 
problem of Transit-Time damping, but common to all the aforementioned 
acceleration mechanisms. 
The isotropization in case of particle acceleration by a sub-Dreicer 
DC field has been investigated by \citet{moghaddam90}. Assuming
quasi-linear diffusion and a wave spectrum due to the runaway
electrons \citep{gandy83}, they found that the actual 
isotropization mechanism is anomalous Doppler resonance of fast
particles and that a bump in the reduced velocity distribution 
is formed. 

In this paper, the isotropization process of an electron temperature 
anisotropic plasma without further energy input is addressed. 
The free energy  in the  anisotropic electron velocity
distribution may drive an instability. The instability would lead
to wave growth which in turn can scatter the particles in velocity
space. One proposed instability for a electron temperature anisotropic
plasma is the Electron Firehose Instability (EFI) 
\citep{hollwegvoelk70,paesoldbenz99,lihabbal00}. It is a
kinetic version of the MHD firehose instability 
\citep{rosenbluth56, parker58} with a lower threshold in electron 
velocity anisotropy. But is this mechanism actually responsible for the 
isotropization of the electron velocity anisotropy?
The previous investigations of this mechanism were based on 
linear theory, thus stopped before wave growth effects the particle 
distribution. If the EFI actually grows or if another instability 
isotropizes the plasma quicker cannot be answered within the framework 
of linear theory. Also  questions about wave growth saturation are  
beyond reach of linear theory.

Here, the isotropization mechanism of an electron temperature anisotropy 
is investigated by means of particle-in-cell (PIC) simulations, giving 
access to the non-linear effects on the particle 
distribution as well as the saturation of the instability. 
The parameters for the simulation are chosen to be compatible with
parameters expected under solar flare conditions. While
rough estimates for density, ambient magnetic field and electron 
temperature exist \citep[e.g.][]{miller97}, values of the actual anisotropy
depend on the exact details of the acceleration mechanism.
The anisotropy chosen here is  $T^e_\|/T^e_\perp = 20$ in
accordance with earlier estimates \citep{paesoldbenz99}. 
The simulated plasma corresponds to a population of electrons
accelerated from $3$keV to $25$keV without any instabilities. 

The paper is organized as follows: Section \ref{section:simulation} 
describes the simulation model, followed by a presentation of the
overall development of the isotropization process in
Sect.~\ref{section:development}. The responsible instability is then
identified by comparison with linear theory in
Sect.~\ref{section:linear}. Non-linear processes are discussed in
Sect.~\ref{section:nonlinear}, followed by an investigation of the
energy flow in Sect.~\ref{section:energy}.  A discussion of the 
applicability of result onto real plasmas (Sect.~\ref{section:real}) 
and of the relevance of the described mechanism summarizes the 
results in Sect.~\ref{section:summary}.

\section{Simulation Model}
\label{section:simulation}
The simulations were performed with the fully 3D relativistic
electromagnetic particle-in-cell code {\em par-T} \citep{messmer00}, 
a parallel implementation of TRISTAN \citep{buneman93}. 
As both particle species are expected to play a role in the electron
firehose instability, they both have to be treated kinetically. 
This prevents the use of hybrid simulations which treat one particle
species as a fluid. 

A PIC code  traces the trajectories of a representative number of
particles under the influence of the Newton-Lorentz force in their 
self-consistent electromagnetic fields. The particles can be placed 
anywhere within the computational domain, whereas the field quantities
are located on a regular spatial grid. Subgrid resolution, e.g. in
order to determine  the forces at the particle positions, is obtained
by linear interpolation. For an introduction on PIC see 
e.g.~\citet{birdsall85}.

Time integration for both  particles and  fields is performed by a 
leap-frog scheme. 
The fields are staggered in space \citep{yee66}, allowing e.g. to 
update the B-field by computing the curl of the
E-field, using finite differences. This method has the advantage of being
simple to implement and additionally to keep $\nabla B = 0$, if it was
satisfied initially. In oder to update the E-field, the current
density has to be determined in addition to the curl of B. 
Instead of solving the Poission equation explicitly, the herein used
code applies a charge conserving current deposition 
algorithm \citep{villasenor92}. This scheme allows to update the
E-field entirely from information within a few grid
cells. Additionally it keeps $\nabla E = \rho$ satisfied, 
if it was satisfied initially, down to machine accuracy. 

The size of the simulation box is 
$L_x \times L_y \times L_z = 8 \Delta\times 8\Delta \times 768\Delta$,
where $L_j$ is the length of the system in dimension $j$, 
$\Delta = 0.13 c / \omega_e$ represents the cell size, $\omega_e$
is the electron plasma frequency and $c$ is the speed of light.  
Periodic boundary conditions in all three dimensions are applied on
both particles and fields. The total particle number is 
$2 \cdot n_p = 3.17 \cdot 10^6$ 
particles, an average of 32 electron-proton pairs per cell. 
The particles are placed uniformly within the computational domain. 
The simulation time step is  $\omega_{e} \Delta t= 2\pi/100$. 
Along the $z$-axis, the external magnetic field $B_0$ is applied,
leading to a ratio $\Omega_e/\omega_e=0.13$, where $\Omega_e$ is the
electron cyclotron frequency. Assuming a particle 
density of  $n=5 \cdot 10^{10}{\rm cm}^{-3}$, the external magnetic 
field is $B_0 = 91$~Gauss.
As the isotropization is expected to take place on
proton time scales, an artificial proton/electron-mass ratio  $m_p/m_e = 49$ 
is chosen, a tribute to computing time.   
The proton plasma frequency is therefore
$\omega_p / \omega_e = 0.14$ and the proton cyclotron frequency 
$\Omega_p / \omega_e = 0.0026$.
The initial electric field is $E_0 = 0$ everywhere within the computational 
domain by placing electrons and protons at the same positions. 

Initially, the electron temperature parallel to the external field is 
$T_\|^e=2.3\cdot 10^8$ K  while the perpendicular electron temperature
$T_\perp^e$ is identical to the isotropic proton temperature $T^p= 1.2
\cdot 10^7$ K. The electron temperature ratio is therefore 
$T_\|^e / T_\perp^e \approx 20$. The corresponding thermal velocities
are for electrons $v_\|^e/c = 0.2$ and $v_\perp^e/c =  0.045$. 
The Debye lengths are $\lambda_{D,\|} = 1.53\Delta$ and
$\lambda_{D,\perp}=0.34\Delta$. 
 
By choosing a rod shaped geometry, wave numbers at oblique
angles are limited to values $k > k_\perp^{\min} = k_\|^{\min}
L_z / L_x = 42 \omega_p/c$, where $k_\|^{\min}$ is the minimum wave
number in parallel direction. The simulation can therefore be
considered as a 1D simulation. 

The instability is expected to grow on timescales of proton cyclotron
periods. This makes it necessary to run the simulation at least for
several proton cyclotron periods. On the other hand, the free energy is
carried by the electrons, requiring resolution of the electron plasma 
frequency. Even with an artificially low mass ratio, this leads to
several thousand timesteps per inverse proton cyclotron frequency. 
To overcome the enormous computing effort, a parallelized simulation code
was applied. Ideally, such a code requires a computing time 
$\tau_p = \tau /N_p$, where $N_p$ is the number of processors, as compared
to  $\tau$, the time required by  the sequential code. In a more realistic 
model, the speedup $S = \tau / \tau_p < N_p$, mainly due to the
overhead introduced by the communication among the  processors. 
However, in case of a particle-in-cell code with charge conserving 
current update \citep{villasenor92} only local information is used 
to solve the field equations. Consequently most of the time is spent
in pushing particles and updating field quantities, letting
communication be of minor influence. The code does therefore not
benefit from a highly sophisticated interconnection network. 
It can easily be run on a cluster of workstations, which are usually 
much cheaper and more available than commercial parallel computers. 
The simulations were therefore performed on a Beowulf-Cluster featuring 48
Pentium-III processors. The total simulation time of 120'000 time steps
requires 24 hours, including diagnostics. At a lower limit of the
parallel  efficiency of $\epsilon \approx 0.8$ on $48$ processors 
\citep{messmer00}, this  corresponds to about $920$ hours of the 
original TRISTAN code on a  single Pentium III processor. 


\section{Overall Development}
\label{section:development}
\begin{figure}
\begin{center}
\resizebox{8.8cm}{!}{
    \includegraphics{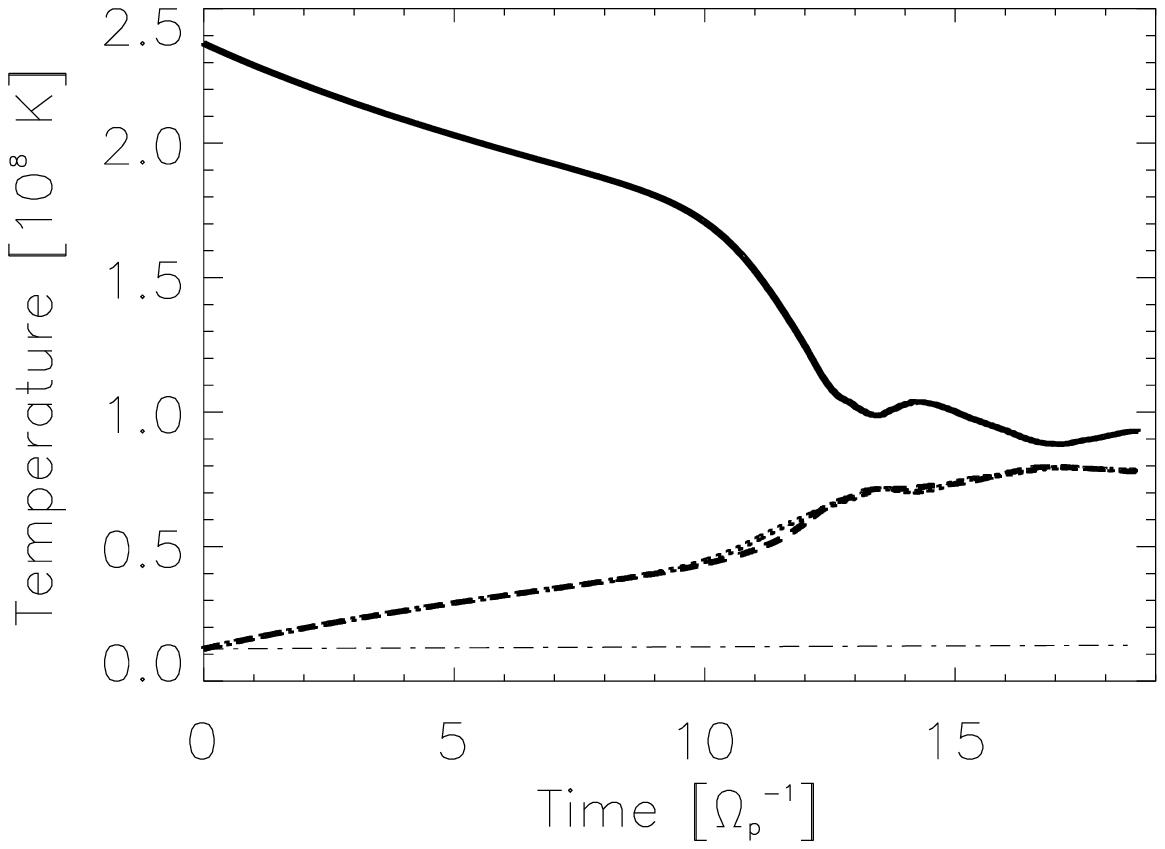}
}
\hfill
\resizebox{8.8cm}{!}{
    \includegraphics{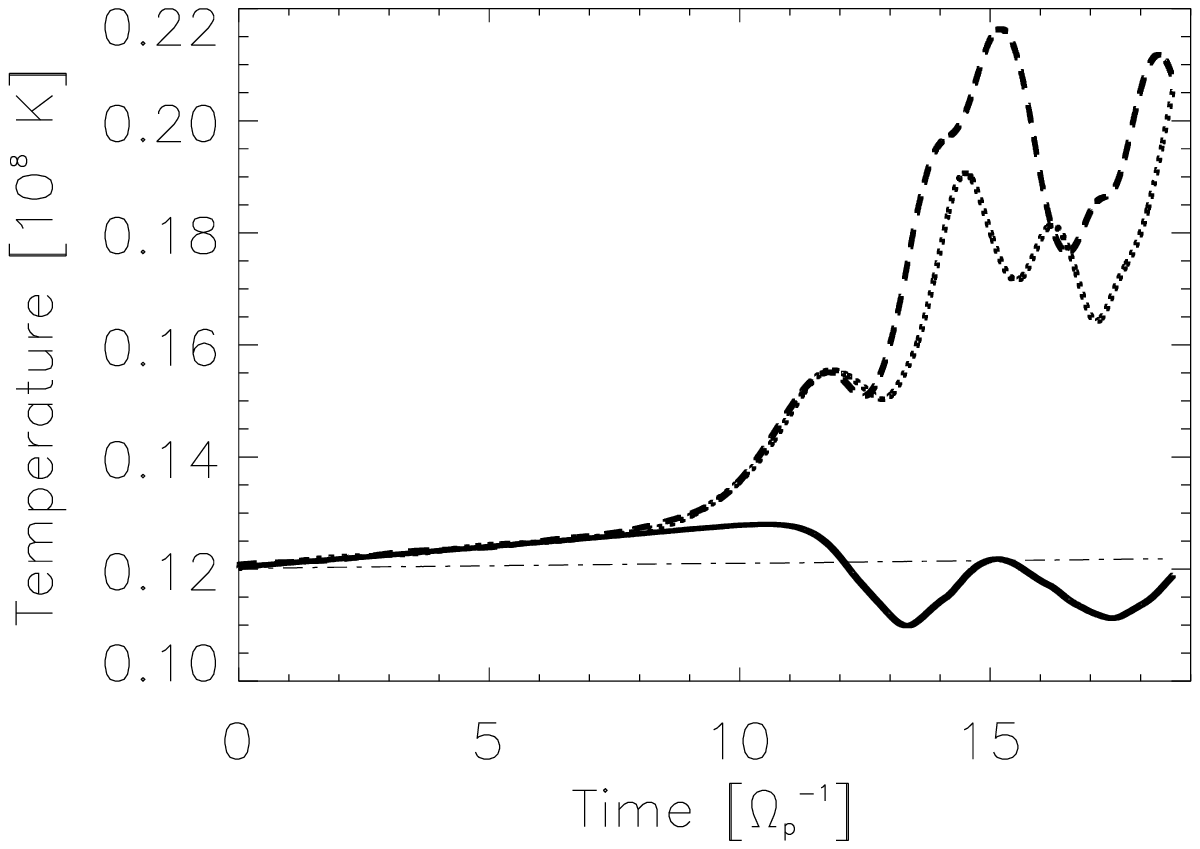}
}
\end{center} 
\caption{Temporal development of the local electron temperatures
(top) and proton temperatures (bottom), compared to the temperature
development in case of an isotropic plasma (thin dash-dotted lines). 
Temperatures are shown along  the three spatial dimensions 
$x, y, z$ (dotted, dashed, solid).  
}
\label{temp}
\end{figure}
Figure~\ref{temp} shows the development of the local temperatures for both
the electrons and the protons during the instability.
As both the electron and the proton velocity distribution may not stay  
Maxwellian throughout the instability, a global temperature is not
defined.  The local temperature is defined as the standard deviation
in one velocity component for a set of particles in a small subdomain
of the computational domain. The temperature of the distribution is 
then defined as the average of all local temperatures.

During the simulated time of $18\Omega_p^{-1}$,  the inital 
electron temperature ratio reduces to  $T_\|^e / T_\perp^e \le 1.2$.
The free energy in the electrons goes partially into the waves, but
mainly into the perpendicular electron temperature.  This will be
shown in Sect.~\ref{section:energy}.
Additionally the protons are slightly heated in all directions. 
At about  $8 \Omega_p^{-1}$, the proton velocity 
distribution is becoming anisotropic with  $T^p_\perp > T^p_\|$.
The oscillations are an indicator of energy transfer between the
protons and the waves.
 
For comparison, the parallel temperature development of an
isotropic plasma is also plotted in Fig.~\ref{temp}. As expected, the
temperature does practically not change throughout the simulated time. This is a
first indicator of energy conservation within the simulation. 

\begin{figure}
\begin{center}
\resizebox{8.8cm}{!}{
        \includegraphics{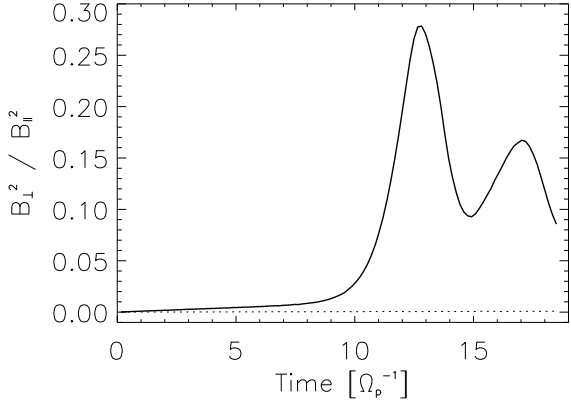}
}
\end{center} 
\caption{Temporal development of $B_\perp^2/B_\|^2$ for the anisotropic
electron velocity distribution with $T_\|^e/T_\perp^e\approx 20$
(solid) and  $T_\perp^e=T^p$,  compared to the development of 
$B_\perp^2/B_\|^2$
for an isotropic electron distribution with $T_\|^e=T_\perp^e=T^p$ (dotted).}
\label{btrans}
\end{figure}
Figure~\ref{btrans} shows the time history of the perpendicular
magnetic field energy  
$B_\perp^2=\int\!\!\!\int\!\!\!\int ( B_x^2 + B_y^2 ) dx\,dy\,dz$. 
The development of $B_\perp^2$ for a simulated
plasma with isotropic temperatures  $T_\|^e=T_\perp^e=T^p$ but
otherwise identical parameters is again included for comparison. 

After a slow increase during $8 \Omega_p^{-1}$, the energy in 
$B_\perp$ increases explosively, saturates at $12.5 \Omega_p^{-1}$ and 
then starts to oscillate, indicating  the growth of transverse 
(electro)magnetic waves. 

\section{Comparison with Linear Theory}
\label{section:linear}

\begin{figure}
\begin{center}
\resizebox{8.8cm}{!}{
      \includegraphics{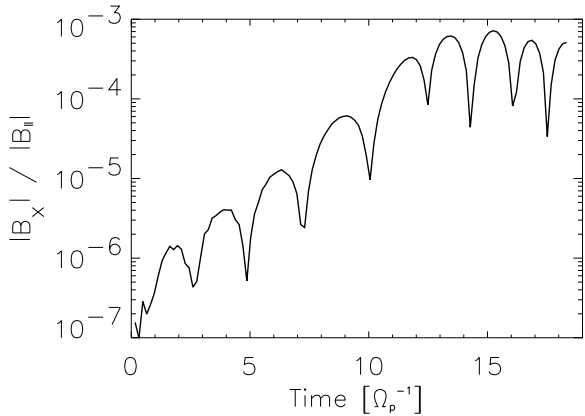}
}
\end{center} 
\caption{Temporal development of $|B_x|$ for the wave number 
$k c / \omega_p =1.32$.}
\label{freq}
\end{figure}
In order to investigate the isotropization further, the excited waves 
have to be identified. 
The wave numbers are determined by Fourier transform of $B_x$, 
one component of the perpendicular B-field.  
Figure~\ref{freq} shows the temporal evolution of the spatial mode 
$kc/\omega_p = 1.32$ propagating parallel to $B_0$.  
After an initial phase of
$\approx 1 \Omega_p^{-1}$, a regular oscillation starts with
exponentially growing amplitude. Close to  $13 \Omega_p^{-1}$, the wave
growth stops and the oscillation continues.   


To identify the wave mode responsible for the growing field $B_\perp$, the
growth rates of  different wave numbers are compared 
to the values obtained by WHAMP \citep{roennmark82}. This program
allows to evaluate the plasma dispersion function numerically for
arbitrary temperature ratios.
 
\begin{figure}
\begin{center}
\resizebox{8.8cm}{!}{
      \includegraphics{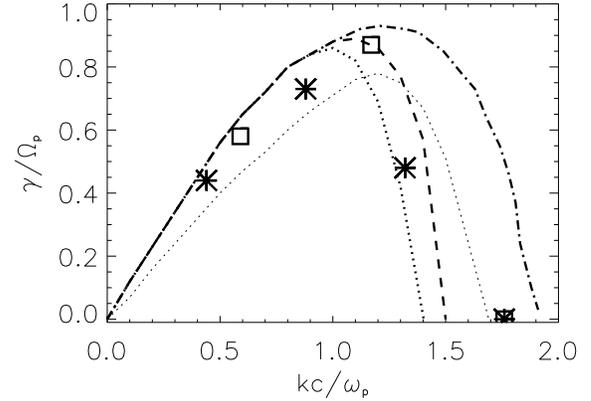}
}
\end{center} 
\caption{Growth rate of $B_x$ determined from simulations with
$m_p/m_e = 49$ (asterisk) and $m_p/m_e = 81$ (boxes) compared to results 
obtained from linear theory for $m_p/m_e = [49, 81, 1836]$ 
(dotted, dashed, dash-dotted) for the initial electron temperature
ratio $T^e_\|/T^e_\perp\approx 20$. In addition, the growth rate for 
$T^e_\|/T^e_\perp\approx 6$ and $m_p/m_e=49$ is shown (dotted, thin).}
\label{growth}
\end{figure}

Figure~\ref{growth} shows a comparison of the growth rates between
linear theory and PIC simulation. The linear theory results are
determined for the initial plasma parameters. 
The growth rates of the PIC simulation are measured by a linear 
fit in the time series of spatial Fourier transforms of $B_x$ in the
time interval $2-8\Omega_p^{-1}$.
The three growing waves and the suppressed growth at $k c/\omega_p =
1.76$ are well in accordance with linear theory. However, the growth
rates for $kc/\omega_p=[0.44,0.88]$ are too low, whereas the one for 
$kc/\omega_p=1.32$ is too high. This will be investigated in 
Sect.~\ref{section:nonlinear_temp}.

In order to check the influence of the artificial mass ratio, the
growth rates are determined for an increased mass ratio of
$m_p/m_e=81$  and otherwise identical plasma parameters (see
Fig.~\ref{growth}).
Due to the 
larger mass ratio and the fixed simulation box size of $100
c/\omega_e$, the resolution in wave numbers decreases. Again, the
waves grow in accordance with linear theory, but the determined growth
rates are lower than expected for $kc/\omega_p=0.59$ and a bit larger
for $kc/\omega_p=1.18$.   
For comparison, the growth rates for the real 
mass ratio  $m_p/m_e=1836$ is included in Fig.~\ref{growth}. The
hypothetical light weight protons inhibit wave growth at large k.

\begin{figure}
\begin{center}
\resizebox{8.8cm}{!}{
      \includegraphics{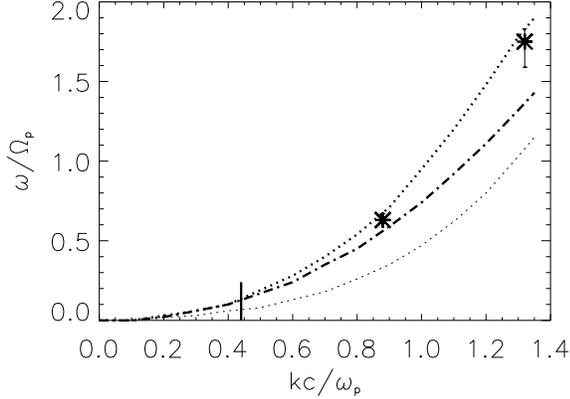}
}
\end{center} 
\caption{Comparison of the dispersion relations obtained by the
simulation (asterisk) and by WHAMP (dashed). For comparison, the
dispersion relation for the real mass ratio $m_p/m_e=1836$ is included
(dash-dotted).}
\label{dispersion}
\end{figure}

Figure~\ref{dispersion} shows a comparison of the dispersion relation 
determined both by WHAMP and from the simulation.
Due to the limited resolution in k-space, only 3 samples lie within the
range of non-zero growth (see Fig.~\ref{growth}). 
According to linear theory, the period of the sample with smallest 
wave number, $kc/\omega_p =0.44$,  is $45 \Omega_p^{-1}$ and  thus
much longer than the saturation time of the instability. The
frequency can therefore not be measured in the simulation
and only an upper bound can be given.
For the other two samples, the frequencies are  determined by 
removing the exponential growth and measuring the time
between the first two amplitude minima, thus the first half period. 
Like the growth rates, the wave frequencies agree well with the values
determined from linear theory. However, they are both too low. 

\begin{table}
\centering
\caption[]{Comparison of growth rates $[\gamma_{0.44},\gamma_{0.88},
\gamma_{1.32}]$ between simulation (bold) and linear theory 
at $kc/\omega_p = [0.44, 0.88, 1.32]$ and wave number of maximum
growth $k_m$  for different electron temperature ratios $T^e_\|/T^e_\perp$.
The other
plasma parameters are identical to Sect.~\ref{section:simulation}.}
\begin{tabular}{ccccc}
$T^e_\|/T^e_\perp$ & $\gamma_{0.44}/\Omega_p$ & $\gamma_{0.88}/\Omega_p$ &  
$\gamma_{1.32}/\Omega_p$ & $k_mc/\omega_p$ \\ \hline\\
10 & ${\bf 0.2}\;\;\;0.2$ & ${\bf 0.4}\;\;\;0.4$  & ${\bf 0.3}\;\;\;0.4$ &$1.6$\\
15 & ${\bf 0.2}\;\;\;0.4$ & ${\bf 0.6}\;\;\;0.7$  & ${\bf 0.7}\;\;\;0.7$ &$1.2$\\
20 & ${\bf 0.4}\;\;\;0.4$ & ${\bf 0.7}\;\;\;0.7$  & ${\bf 0.5}\;\;\;0.4$ &$1.0$\\
30 & ${\bf 0.6}\;\;\;0.7$ & ${\bf 0.6}\;\;\;0.8$  & ${\bf 0.0}\;\;\;0.0$ &$0.7$\\
\end{tabular}
\label{aniso_table}
\end{table}

Additional simulations with varying initial temperature ratio have 
been carried out to test the agreement  with linear theory. 
Table~\ref{aniso_table} shows the growth rates for
varying temperature ratios $T^e_\|/T^e_\perp$, but otherwise identical
parameters as in Sect.~\ref{section:simulation}. A larger anisotropy
leads to a shift of the maximum growth rate towards smaller wave numbers 
(see also Fig.~\ref{growth}).

The good agreement of both dispersion relation and growth rate
indicates that the instability under investigation is indeed the Electron
Firehose Instability.

\section{Non-Linear Effects}
\label{section:nonlinear}
\subsection{Decreasing Temperature Anisotropy}
\label{section:nonlinear_temp}
The previous section showed that the rough estimates for the growth rates 
and the wave frequencies agree well with linear theory.  
The discrepancies can be qualitatively understood as an effect of the 
temperature anisotropy reduction: 
As mentioned in the last section, the growth rates are determined by
linear fit between
$2 - 8 \Omega_p^{-1}$. They are therefore average growth 
rates over the whole time interval. 
However, during this time the temperature ratio is reduced to a value 
$T^e_\|/T^e_\perp \approx 6$ (see Fig.~\ref{growth}). This has an 
influence on the growth rates of the EFI, shown in Fig.~\ref{growth}: 
For lower anisotropy, the maximum 
growth rate is shifted towards larger wave numbers.
For the reduced anisotropy, the growth rates for $kc/\omega_p < 1.1$
are lower and for $kc/\omega_p > 1.1$ are larger than for the initial
anisotropy. Averaging the growth rates at $kc/\omega_p<1.1$ thus
underestimates the initial growth rates, whereas for larger wave numbers
the growth rates are overestimated. This agrees with the
measurements in Fig.~\ref{growth} and Tab.~\ref{aniso_table}. 

\begin{figure}
\begin{center}
\resizebox{8.8cm}{!}{
      \includegraphics{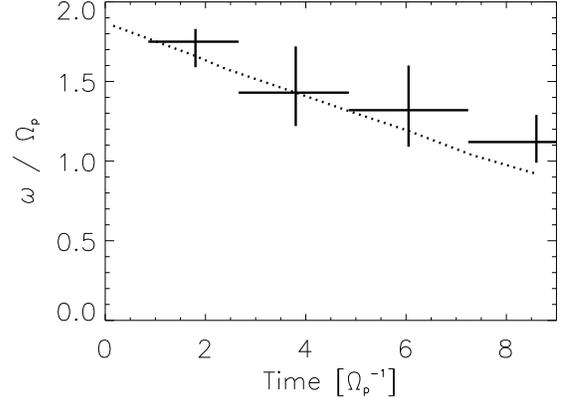}
}
\end{center} 
\caption{Temporal development of the wave frequency for
$kc/\omega_p=1.32$ compared to the frequencies expected from linear
theory according to the instantaneous temperature anisotropy.}
\label{omega_decay}
\end{figure}

The reduction of the anisotropy has not only an effect on the growth
rate, but also on the real part of the wave frequency. 
Fig.~\ref{dispersion} includes the dispersion relation for the
reduced temperature ratio $T^e_\|/T^e_\perp\approx6$, showing a frequency
decrease for all wave numbers. Measuring the time between two minima
in spectral power will only  yield an average frequency. 

Figure~\ref{omega_decay} shows the development of the average
frequency for the wave number $kc/\omega_p = 1.32$ (see also
Fig.~\ref{freq}). The
horizontal bars indicate the time interval between two amplitude
minima.  
The errorbars in frequency represent the uncertainty in determing the 
time of the minimum amplitude.  The average frequencies agree
well with the frequencies expected from the linear theory for the
instantaneous temperature anisotropies. 

\subsection{Saturation}
Based on linear theory, \citet{hollwegvoelk70} give an approximate 
instability criterion for growth of a left hand polarized
electro-magnetic wave in a temperature anisotropic plasma, 
\begin{equation}
 1 - \beta_\|^e A_e < 0
\label{eqinstability}
\end{equation}
where $\beta_\|^e= 2 (\omega_e / \Omega_e)^2 \cdot  (v_\|^e / c)^2$ is the
ratio of thermal to magnetic pressure,  and 
$A_e = 1 - T_\perp^e /  T_\|^e$ is the
electron temperature anisotropy. This criterion is strictly valid only
for small anisotropies $|1 - \beta_\|^e A_e| \ll 1$. What happens in case of 
a larger anisotropy?
The threshold is identical to the MHD firehose instability criterion in
case of an isotropic proton velocity  distribution. The simple physical 
interpretation is that the centrifugal force exerted by the electrons 
moving along a disturbed magnetic field line, destabilizes the wave. If
the anisotropy is not large enough, the magnetic field stress and the
thermal pressure perpendicular to the field line act as restoring 
forces. If the velocity anisotropy is large and the magnetic field is
small enough, the restoring force is too weak and thus the wave
destabilizes.

\begin{figure}
\begin{center}
\resizebox{8.8cm}{!}{
    \includegraphics{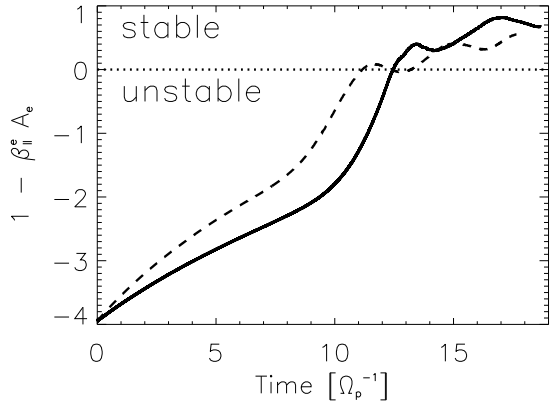}
}
\end{center} 
\caption{Temporal development of $1-\beta^e_\| A_e$ for $m_p/m_e=49$
(solid) and $m_p/m_e=81$ (dashed).  The dotted line 
represents the instability criterion given by Eq.~(\ref{eqinstability}).}
\label{instab}
\end{figure}

Figure~\ref{instab} shows the temporal development of the anisotropy and 
instability criterion given in Eq.~(\ref{eqinstability}), assuming a 
constant background magnetic field. The initial setup lies
within the unstable region, triggering the instability. As soon as the
stable region is reached at time $12 \Omega_p^{-1}$, wave growth
stops. The instability
criterion then states, that the anisotropy in the electron velocity
distribution has mostly been eroded.

\subsection{Electron Velocity Distribution}
\label{section:electron_velocity_distribution}
An advantage of self-consistent kinetic simulations is the possibility to
investigate effects on the particle distributions due to wave growth. 

\begin{figure*}
\begin{center}
\resizebox{8.8cm}{!}{
        \includegraphics{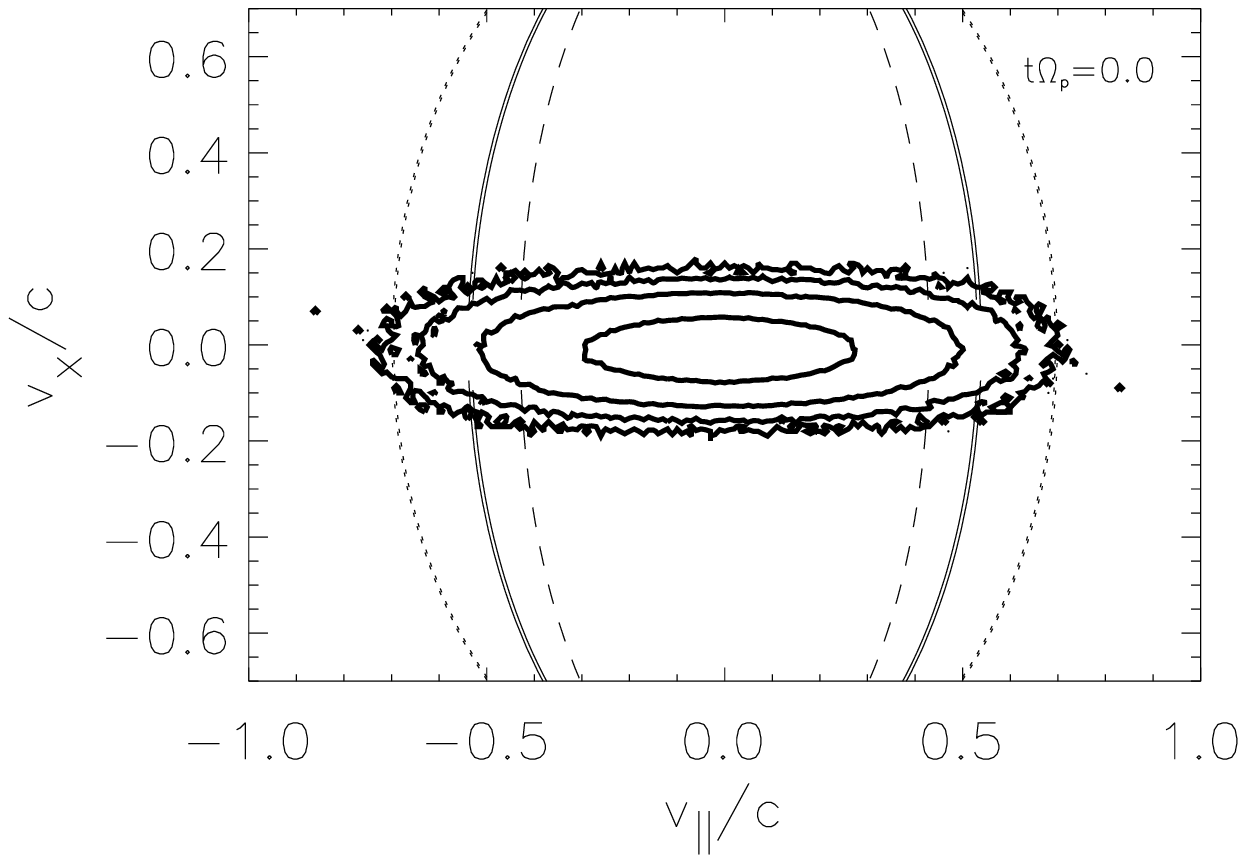}
}
\hfill
\resizebox{8.8cm}{!}{
        \includegraphics{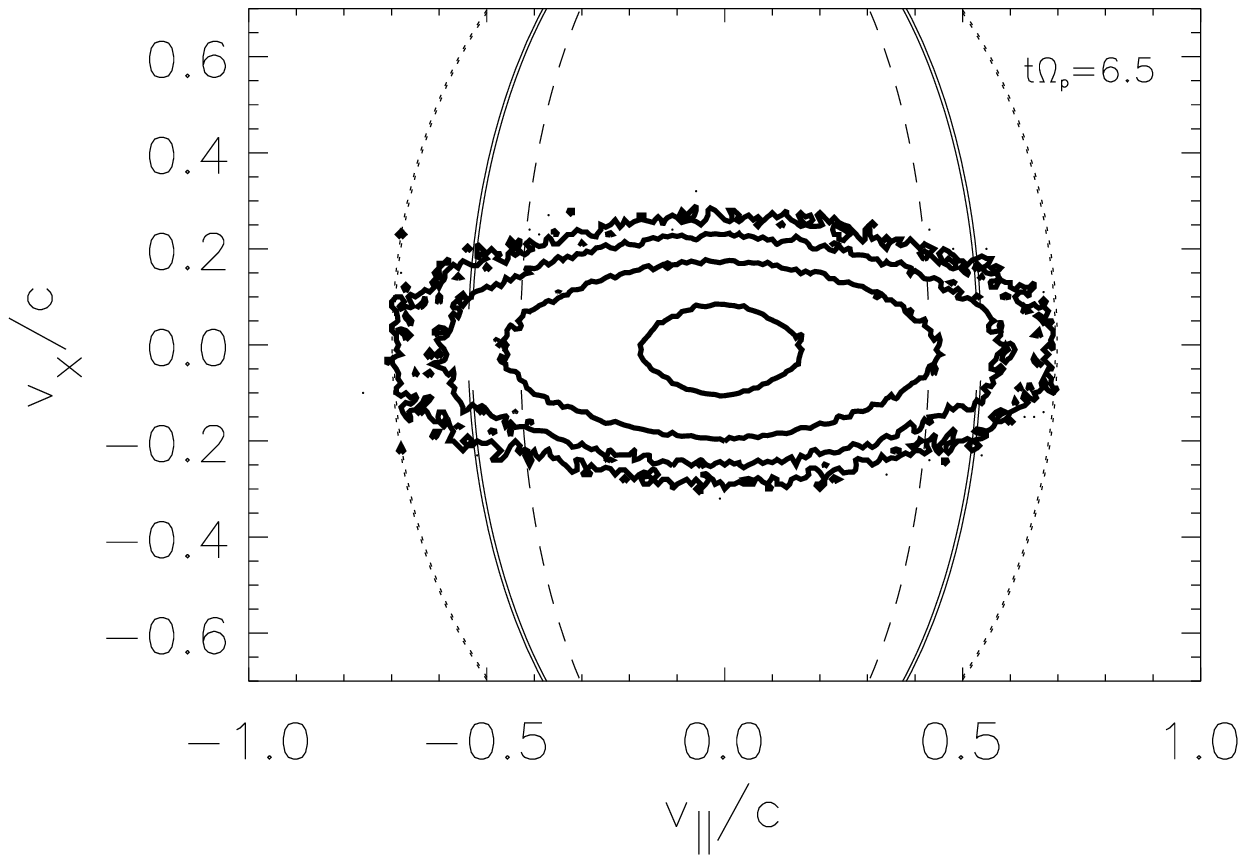}
}
\hfill
\resizebox{8.8cm}{!}{
        \includegraphics{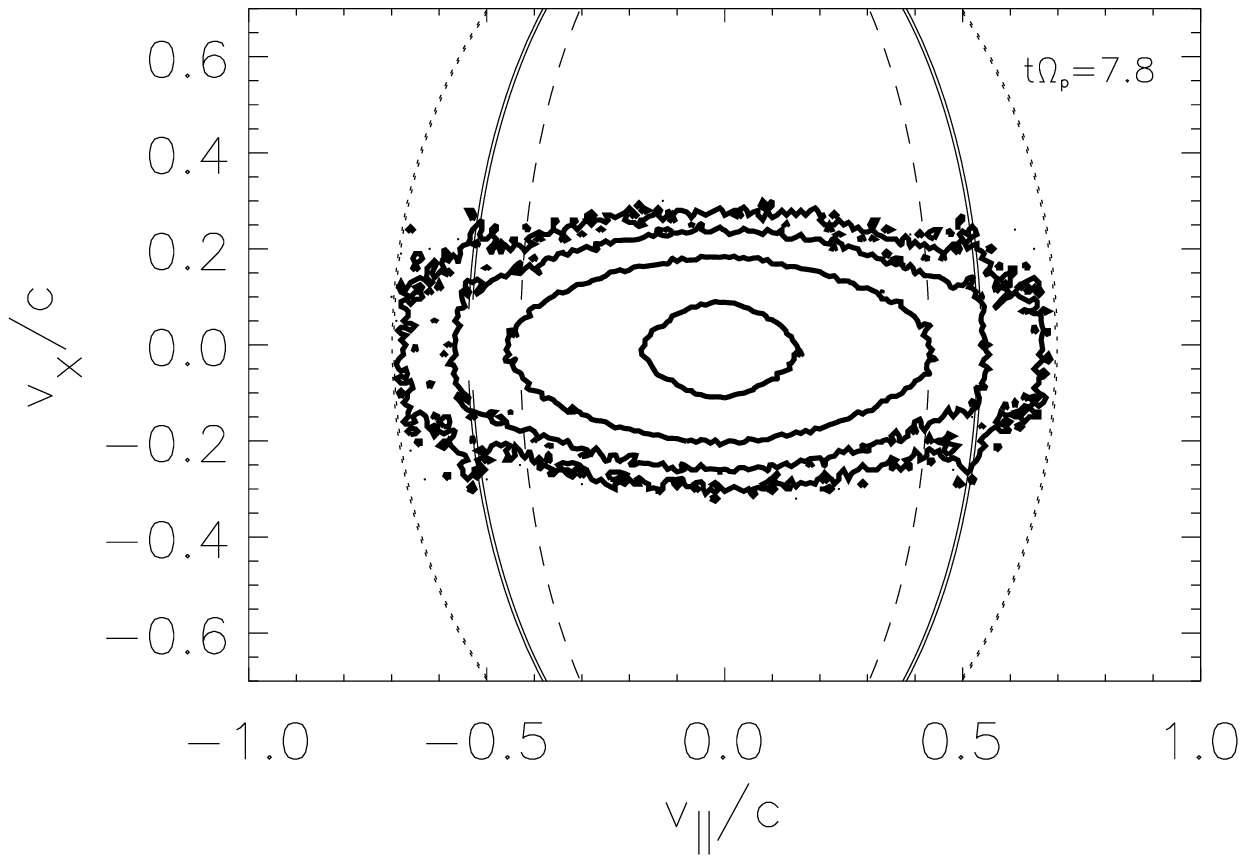}
}
\hfill
\resizebox{8.8cm}{!}{
        \includegraphics{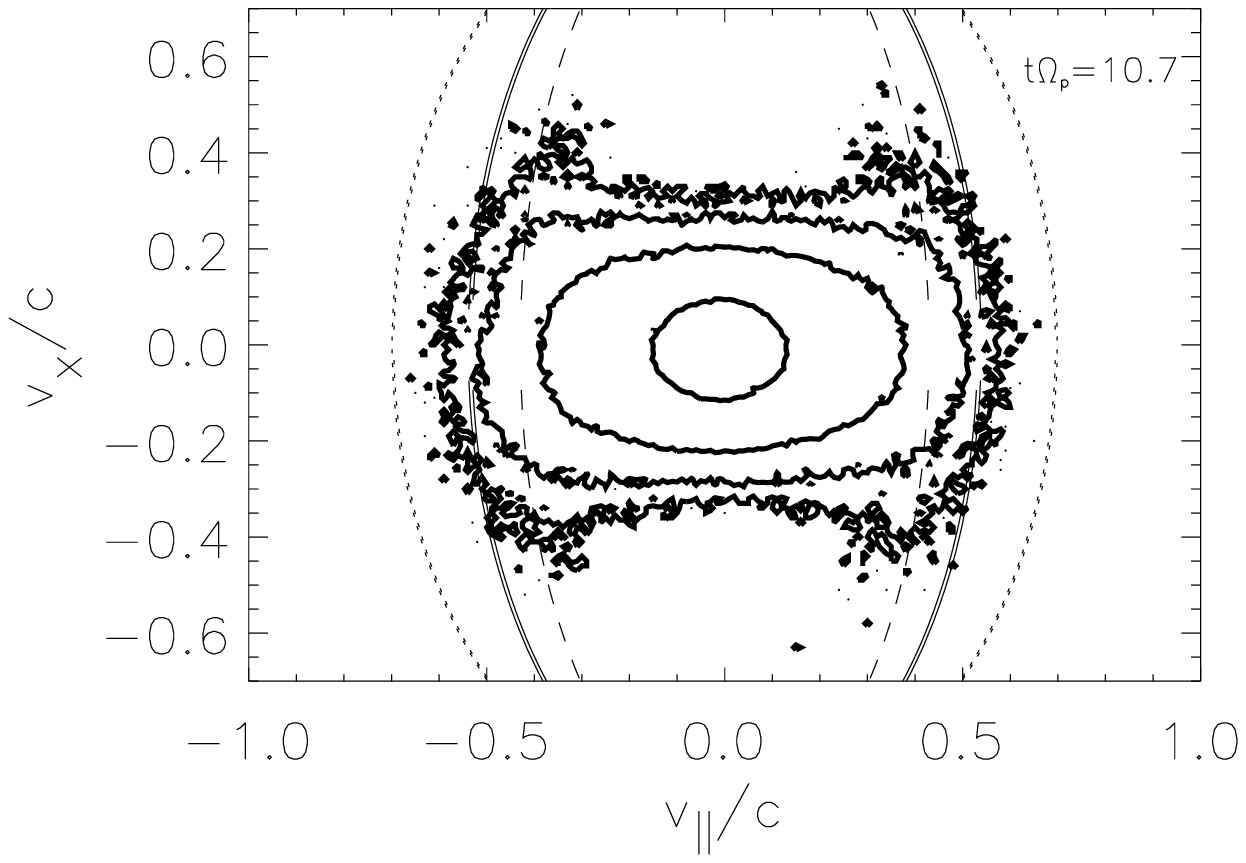}
}
\hfill
\resizebox{8.8cm}{!}{
        \includegraphics{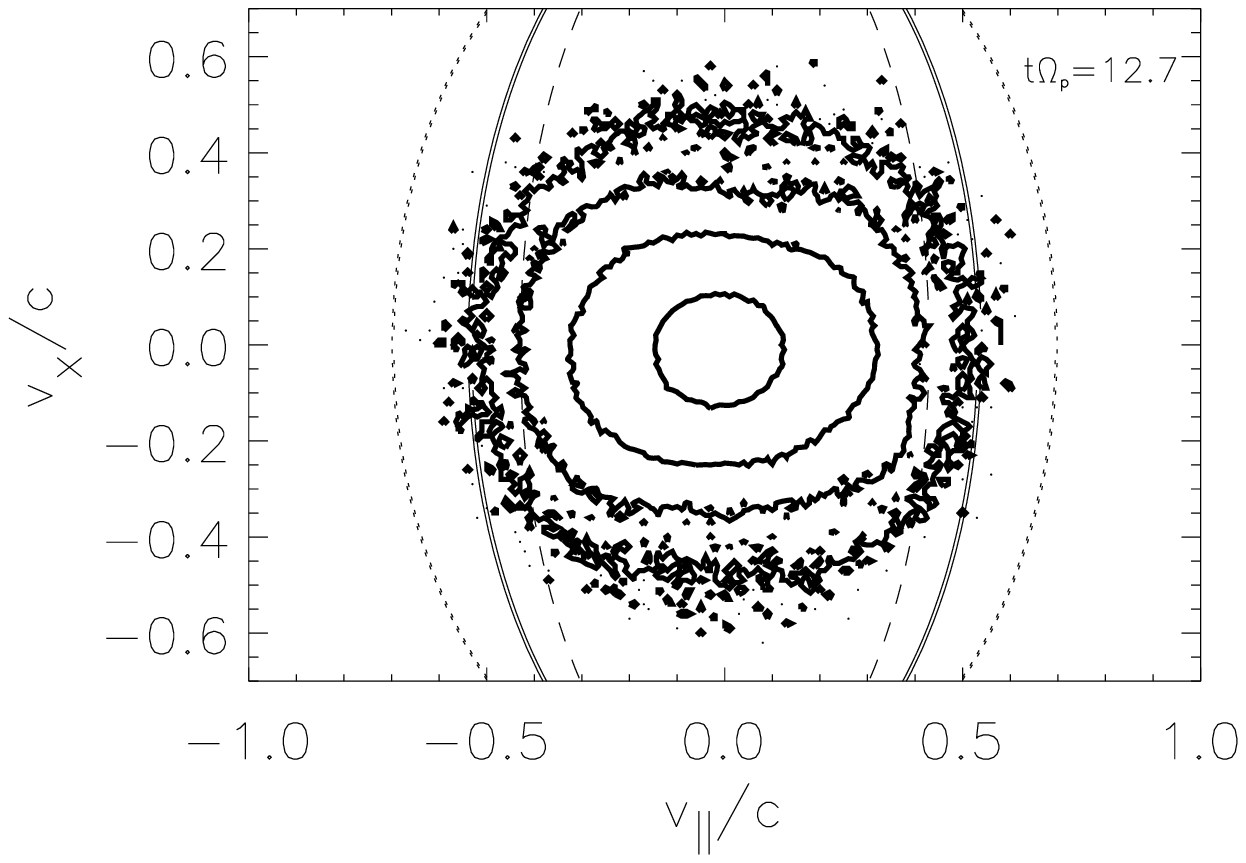}
}
\hfill
\resizebox{8.8cm}{!}{
        \includegraphics{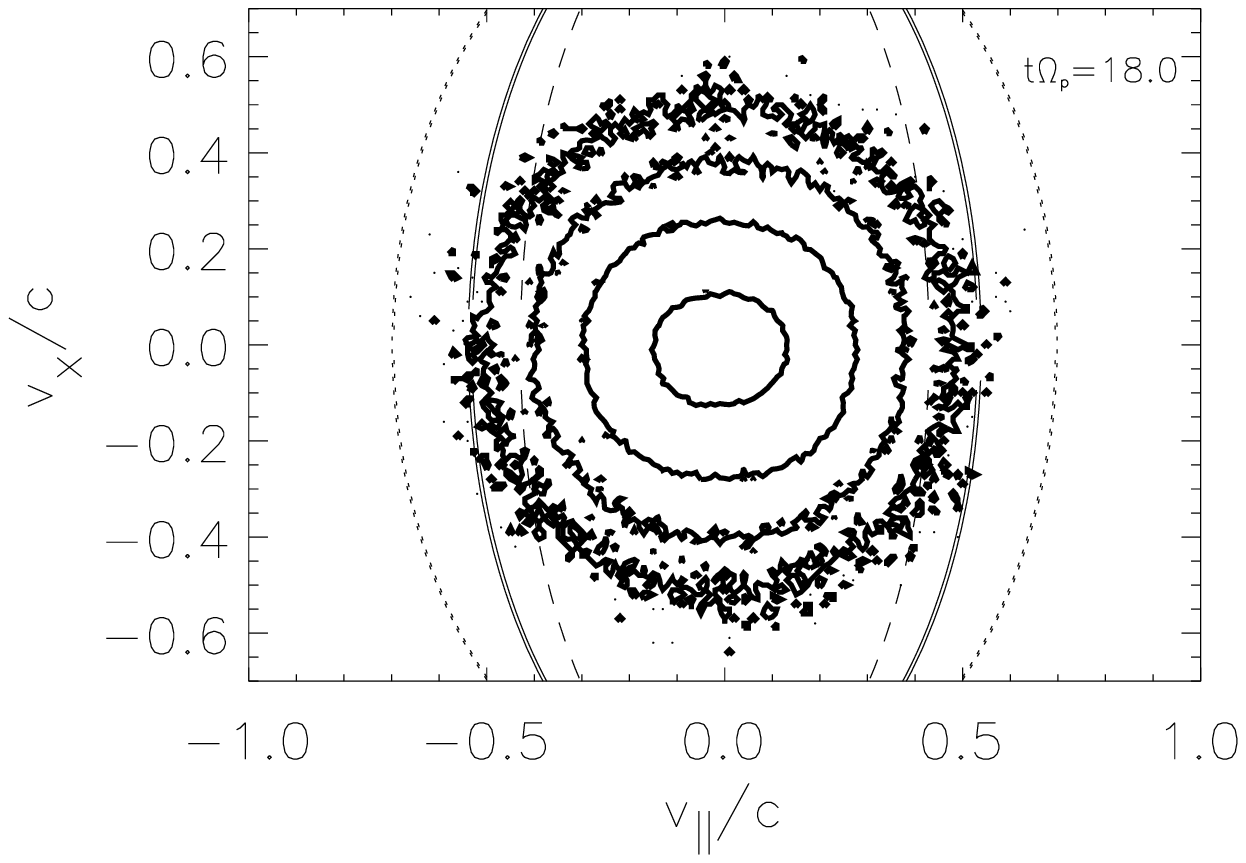}
}
\end{center}
\caption{Projection of the  electron velocity distribution along $v_y$ 
at times $t = [0, 6.5, 7.8,10.7,12.7, 18.0] \Omega_p^{-1}$ 
(left to right, top to bottom). The contour levels are at 
$[1, 10, 100, 1000]$ particles. The ambient magnetic field points from
left to right. 
Overplotted are the resonance curves for the three modes
$kc/\omega_p=[0.89, 1.32, 1.76]$ (dotted, solid, dashed). The aspect
ratio is chosen such that circular contour lines correspond to a
thermal distribution.}
\label{distr_electrons}
\end{figure*}
Figure~\ref{distr_electrons} shows the electron velocity distributions 
at times $t=[0, 6.5, 7.8, 10.7,12.7, 18.0] \Omega_p^{-1}$. 
It demonstrates that the initial velocity  distribution 
isotropizes in the course of the simulated $18\Omega_p^{-1}$. 
This is a result of the EFI growing at the same time scales as the 
isotropization takes place.  
A closer look at the velocity distribution reveals two distinct features: 
on one hand, for the bulk of the electron velocity 
distribution,  $v_\perp$ increases at small $v_\|$. On the other
hand, particles with $|v_\|| \gg |v_\|^e|$ are ejected from the distribution 
to larger pitch angles in narrow velocity ranges 
(see Fig.~\ref{distr_electrons}, $t\Omega_p=7.8$).  
The first feature is a sign of non-resonant wave-particle interaction,
whereas the second is a sign of resonant wave-particle interaction.

\subsubsection{Non-Resonant Wave-Particle Interaction}
According to Hollweg and V\"olk (1970), most of the electrons gain 
perpendicular 
momentum by a non-resonant wave-particle interaction. The effect of
such an interaction is pitch-angle scattering of electrons with no
preferred $v_\|$, increasing
$|v_\perp|$ at the expense of $|v_\||$ by randomly fluctuating
wave fields. The non-resonant pitch-angle scattering
can be seen in Fig.~\ref{distr_electrons} before 
$t\approx 7.8\Omega_p^{-1}$, where the perpendicular velocity
increases at all $v_\|$ concurrently. But is it actually pitch-angle
scattering or heating in perpendicular direction? 
 

\begin{figure}
\begin{center}
\resizebox{8.8cm}{!}{
    \includegraphics{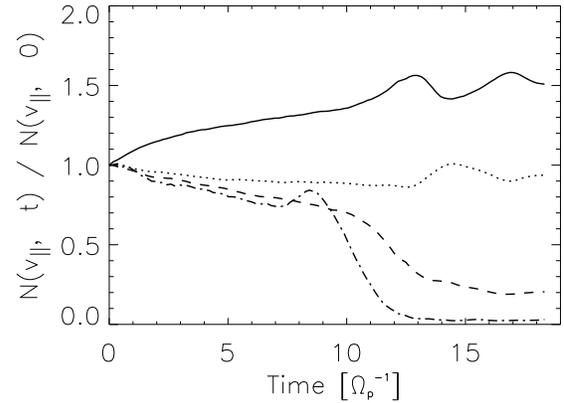}
}
\end{center} 
\caption{Number of particles $N$ in a velocity range of $\Delta v_\|/c =0.04$  
centered at a fixed $v_\|/c = [0, 0.16, 0.30, 0.50]$ 
[solid, dotted, dashed, dash-dotted] 
as a function of time.}
\label{num}
\end{figure}

Figure \ref{num} shows the temporal development of the electron number
density $N$ in a velocity range of $\Delta v_\|/c = 0.04$ centered at
a fixed  $v_\| / c = [0, 0.16, 0.30, 0.50]$. The particle numbers are normalized
to the particle numbers at $t=0$. 
The particle number for $v_\|/c = 0.16\approx v_\|^e/c$ remains about constant
throughout the growth of the instability.
The particle number at low velocities ($v_\|/c = 0$)  increases with time,
while the particle numbers for $v_\|/c \gg v_\|^e$ decrease. 
 At $v_\|/c = 0.50$  an
increase of particles is visible at $8 \Omega_p^{-1}$ due to the particles
resonant with the wave. This will be discussed in 
Sect.~\ref{section:resonant}. 

The increase of particle number at low $|v_\||$ and the decrease at high
$|v_\||$ indicates that the particles are not simply heated in
perpendicular direction, but that they are scattered towards lower
parallel velocities. 
How is the magnetic field built up?
The scattering of a single particle due to randomly fluctuating fields
corresponds to an acceleration, inducing a magnetic field
pulse,  which propagates as 
a wave. Most of these waves are not linear modes of the given plasma 
and are therefore highly damped. However, some of these waves
correspond to the linear eigenmodes of the
plasma with positive growth rates. After a certain build-up time, the 
amplitude of the induced wave is large enough to influence the
particle velocity distribution. In the present simulation, this 
happens at about $7.5\Omega_p^{-1}$. At this time, fast electrons 
start to  gyroresonate with the growing wave.

%


\subsubsection{Resonant Wave-Particle Interaction}
\label{section:resonant}
In presence of an external magnetic field, anomalous gyroresonance of the 
electrons in the form
\begin{equation}
\omega - k v_\| = - s\frac{\Omega_e}{\gamma}
\end{equation}
can be expected, where $\omega$, and $k$ are frequency and wave number of the 
resonant wave, $v_\|$ is the parallel velocity of the resonant
electrons, $s$ is the harmonic resonance number,  $\Omega_e$ the 
non-relativistic electron cyclotron frequency due to the external magnetic 
field and $\gamma$ the  Lorentz factor of the particle. 

The frequencies for the fastest growing modes are determined by linear
theory (see Sect. \ref{section:linear}). 
The relativistic gyroresonance condition can be solved numerically for an 
electron with parallel speed $\beta = v_\|/c $,
\begin{equation}
\Phi = \kappa \beta \mu + \frac{s}{\gamma}
\label{eqresonance}
\end{equation}
where $\Phi = \omega / \Omega_e$ and $\kappa = kc/\Omega_e$ 
\citep{steinackermiller92}. For a
detailed discussion of gyroresonance see e.g. \citet{benz93}.

The resulting  resonance curves for the growing waves due to
the EFI, $kc/\omega_p = [0.88, 1.32, 1.76]$ are plotted in 
Fig.~\ref{distr_electrons}. 
Although the wave at $kc/\omega_p=1.76$
does not grow initially, it becomes unstable as soon as the
temperature ratio is small enough. This is the case at about  
$8.3\Omega_p^{-1}$. 
The resonance curves are also affected by the reduction in temperature
anisotropy, but the effect is small. 

Up to $ \approx 6.5\Omega_p^{-1}$, no specific features in the
velocity distribution can be seen close to the resonance
curves. 
$B_\perp$ is growing, but it is not yet strong enough to affect 
the electron velocity distribution. 
At $\approx 7.8\Omega_p^{-1}$, particles with
velocities close to the resonance velocity of the fastest growing wave
$kc/ \omega_p = 1.32$ are in resonance and as a consequence, are
scattered towards higher perpendicular velocities. 
According to linear theory, the growing wave is a left hand polarized 
electromagnetic wave. 
The  interaction with the electrons is therefore anomalous
Doppler resonance. 
At a later stage, the wave amplitude of the slower growing waves can
also interact with the electrons, scattering them to even lower $v_\|$
and higher $v_\perp$.

%

Only the high velocity tails of the distribution are
affected  by the anomalous Doppler resonance. The major fraction of
the electron particle  distribution gains perpendicular velocity by
the non-resonant interaction. 

\begin{figure}
\begin{center}
\resizebox{8.8cm}{!}{
    \includegraphics{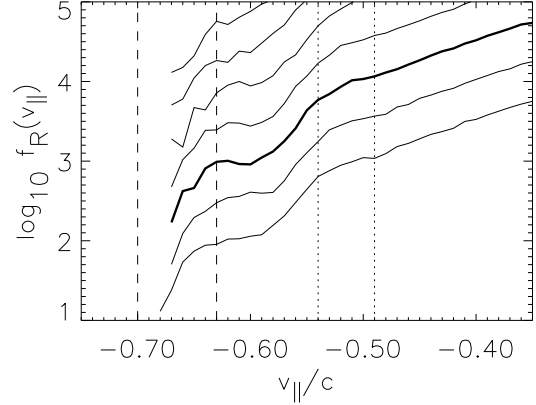}
}
\end{center} 
\caption{Evolution of the reduced electron velocity distribution
$f_R(v_\|)$ between $t=8.11 \Omega_p^{-1}$ (bottom) and 
$t=9.03\Omega_p^{-1}$ (top) at intervals of $0.16\Omega_p^{-1}$. 
The bold line indicates the distribution at $t=8.43\Omega_p^{-1}$ 
with a clearly visible bump. The vertical lines represent the 
resonant velocities $v_\|$ and $0< v_\perp/c<0.4$ for the wave numbers 
$kc/\omega_p = 0.88$ (dashed) and $kc/\omega_p = 1.32$ (dotted).}
\label{beam}
\end{figure}

During the phase of anomalous Doppler resonance, the reduced electron
velocity distribution can generate a bump on the velocity distribution
tail. Figure~\ref{beam} shows the temporal development of the reduced
electron velocity distribution parallel to the external magnetic
field, $f_R(v_\|, t) = \int f(v_\perp, v_\|, t)\,dv_\perp$, where
$f(v_\perp, v_\|, t)$ is the electron velocity distribution at time
$t$,  between times $8.11\Omega_p^{-1}$ and $9.08\Omega_p^{-1}$ at
intervals of $0.16\Omega_p^{-1}$ for
high velocities. Throughout this time interval, a plateau persists
in the distribution between the two resonance velocities for
$kc/\omega_p = [0.88, 1.32]$. 
 At the time $8.43\Omega_p^{-1}$, a clear bump between $v_\|/c=-0.62$ and
$v_\|/c=-0.66$ is visible in the reduced electron velocity
distribution. Due to the symmetry of the problem, a similar plateau
with bump exists for the positive velocities $0.62 < v_\|/c < 0.66$. 
 


A bump on the reduced velocity  distribution leads to the 
bump-on-tail instability, generating Langmuir waves. 
The wave numbers, $k_L$, of the Langmuir waves
generated by particles with velocity $v_\|$ can be estimated by 
combining the Langmuir dispersion relation with the \v{C}erenkov resonance 
condition, 
\begin{equation}
 k_L^2 = \frac{\omega_p^2+\omega_e^2}{v_\|^2 - 3 v_e^2}.
\end{equation}

\begin{figure}
\begin{center}
\resizebox{8.8cm}{!}{
    \includegraphics{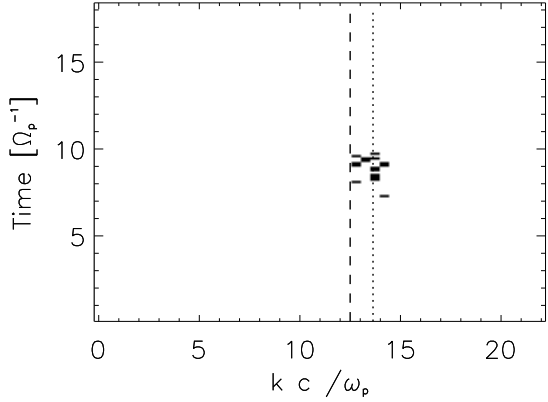}
}
\end{center} 
\caption{Temporal evolution of the parallel electric field amplitude
$E_z$ for different wave numbers. Dark corresponds to locations where
the amplitude exceeds the average field amplitude by $4\sigma$. 
Overplotted are the wave numbers for Langmuir waves generated by a 
parallel propagating weak beam with 
$v_\|/c=0.62$ (dotted) and  $v_\|/c = 0.66$ (dashed).}
\label{eztime}
\end{figure}

Figure~\ref{eztime} shows the temporal evolution of the parallel 
electric field amplitude $E_z$ for different wave numbers. Only field
amplitudes exceeding the mean amplitude by $4\sigma$ are shown,
$\sigma$ being the standard deviation.  
Overplotted  are the Langmuir wave numbers for the beam velocities 
$v_\|/c=0.62$ and 
$v_\|/c=0.66$. The bump-on-tail distribution in Fig.~\ref{beam} 
leads obviously to the emission of Langmuir waves. 

One could expect these longitudinal  waves (L) to couple to the 
transverse EFI waves (T), generating observable transverse radio emission, 
$L + T \rightarrow T$.  
However, the large wave numbers of the Langmuir waves and the 
low frequencies of the EFI waves make it impossible to satisfy the 
parametric equations for observable radio waves. Whether these waves 
can couple to other plasma waves in a more complicated way, yielding
narrowband radio emission,  needs to be investigated further.

\subsection{Proton Velocity Distribution}
\label{section:proton}
\begin{figure}
\begin{center}
\resizebox{8.8cm}{!}{
        \includegraphics{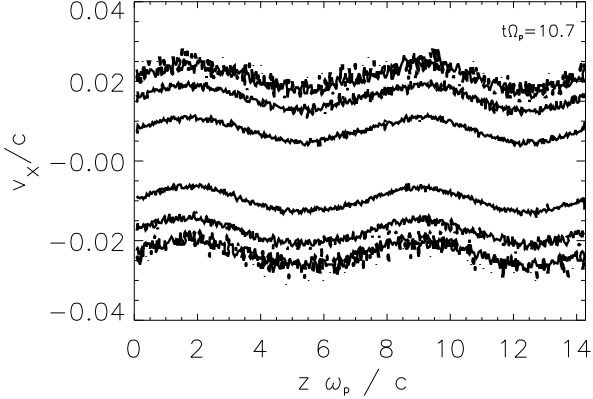}
}
\end{center}
\caption{Projection of the protons in phase space to the plane 
$v_x$ vs. $z$ at $t = 10.7 \Omega_p^{-1}$. 
Contour levels are at $[1, 10, 100]$ particles.
The ambient magnetic field points from left to right. }
\label{phase_ions}
\end{figure}

The initial proton velocity distribution is assumed to be isotropic at a
temperature of $T^p = T_\perp^e$. Although protons do not contribute any
free energy to the instability, they play a significant role in the
course of the instability. As the growing wave due to the EFI is
left hand polarized, the protons come in normal Doppler resonance with the
wave. Unlike electrons, where only the high velocity
tails of the distribution are affected by resonance, the
resonance velocity for the protons lies within the bulk of the proton velocity
distribution. Therefore the protons are expected to absorb energy 
from the wave. The non-collisional electron-proton coupling 
leads to a partial transfer of thermal energy from
the electrons to the protons. This energy transfer was already conjectured by 
\citet{hollwegvoelk70}.

Due to the low thermal velocities of the protons, 
the gyroresonance velocity can be estimated in a
non-relativistic limit, simplifying Eq.~(\ref{eqresonance}) to 
\begin{equation}
 v_\| = \frac{\omega - \Omega_p}{k} = \frac{\omega/\Omega_p - 1}{kc/\omega_p}c_A 
 \label{eq:proton}
\end{equation}
for resonance with an left hand polarized wave, where
$c_A=c\Omega_p/\omega_p$ is the Alfv\'en-speed. The initially fastest
growing waves $(kc/\omega_p; \omega/\Omega_p) = (0.88; 0.68)$
and $(1.32; 1.82)$ have resonance velocities $v_\|/v^p=-1.0$ and
$1.8$. Both waves could be in resonance with some of the protons but
due to the small amplitude of the waves, their influence is small. 

Due to the isotropization of the electron velocity distribution, 
the wave frequencies at fixed wave number change and therefore the proton resonance
velocities have changed to $v_\|/v^p=-2.1$ for $(0.88; 0.33)$ and 
to $v_\|/v^p=0.2$ for $(1.32; 1.07)$ at time $7.3\Omega_p^{-1}$. 
The resonance velocity for  $kc/\omega_p=0.88$ is already far out 
in the proton velocity
distribution and does not affect the proton velocity distribution 
any more. For the wave $(1.32; 1.07)$ on the other hand, a large
fraction of the proton distribution can be in resonance. 
 
Figure~\ref{phase_ions} shows the proton distribution in a projection 
of phase space at $10.7\Omega_p^{-1}$. The spatial wave pattern
corresponds to $kc/\omega_p=0.88$, which is not in resonance with the
protons. The wave with $kc/\omega_p=1.32$ on the other hand is in
resonance with the bulk of the proton velocity distribution, leading
to pitch angle scattering of the protons towards larger $|v_\perp|$
while reducing $|v_\||$. 

\section{Energy and Temperature Development}
\label{section:energy}
The transfer of free energy from the electrons to the protons through 
wave-particle interaction can be observed by considering the relevant energies in
the simulated system. 
\begin{figure}
\begin{center}
\resizebox{8.8cm}{!}{
    \includegraphics{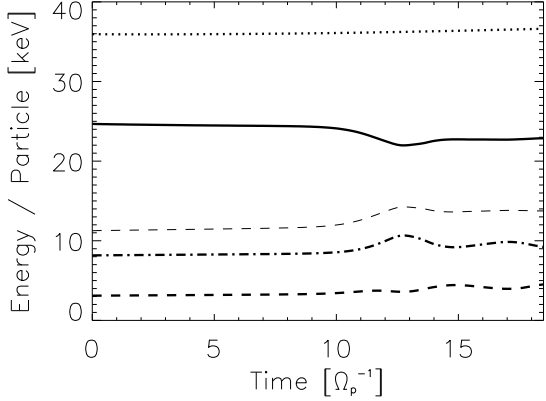}
}

\end{center} 
\caption{Energy development per particle:  Total energy $E^f + E^k_p +
E^k_e$  (dotted), electron kinetic  energy $E^k_e$ (solid), 
electromagnetic field energy $E^f$ (dash-dotted), proton kinetic
energy $E^k_p$ (dashed), and $E^f + E^k_p$ (dashed, thin).
}
\label{total_energy}
\end{figure}

Figure~\ref{total_energy} shows the development of the different
energies per particle, demonstrating the transfer of free electron
energy into waves and from there into proton kinetic energy.
The kinetic energy of species $i$ is given by
\begin{equation} 
E^k_i = \frac{1}{n_p}\sum\limits_{j=1}^{n_p} m_p c^2 (\gamma_j - 1),
\end{equation}
where $\gamma_j$ is the Lorentz-factor of particle $j$. The 
electromagnetic field energy is given by 
\begin{equation}
E^f = \frac{1}{2n_p}\int\!\!\!\int\!\!\!\int ( E^2 + c^2 B^2)
dx\,dy\,dz,
\end{equation} 
where $E^2 = E_x^2+E_y^2+E_z^2$ and $B^2 = B_x^2+B_y^2+B_z^2$.

First it should be noted that throughout the simulated time the 
total energy is conserved to less than $2\%$.

Initially, an electron has on average a total energy of 
$\approx 25$ keV, with $23$ keV in parallel direction and $1$ keV
in each  perpendicular direction. 
The protons are initially isotropic with $1$ keV per proton 
in all directions. 

Up to $\approx 10\Omega_p^{-1}$, the electron kinetic energy remains about
constant. At the same time the parallel electron temperature is
reduced significantly, thus most of the free energy is used to heat
the electrons in perpendicular direction. 
However, some of the electron energy is converted into magnetic field
energy. Due to the high mobility of the electrons, large scale
electric fields do not build up. 
Between $10-12 \Omega_p^{-1}$, energy goes mainly into the magnetic 
field. After saturation, the  electron energy remains constant, 
indicating that the electrons play a passive role in the later 
development. This is also supported by the almost constant $E^f+E^k_p$. 

The protons do not increase their energy significantly up to about
$10\Omega_p^{-1}$. At that time, the bulk of the protons becomes
resonant with the waves, increasing their kinetic energy at the cost 
of the wave. However, due to the resonant character
of the interaction between protons and the waves, the proton  kinetic
energy  can also be fed back into the magnetic field, leading to the 
oscillations  seen after $t\Omega_p \approx 15$. 

\section{Real Plasmas}
\label{section:real}
The isotropization process of an electron temperature anisotropy has been
investigated in case of a plasma with $m_p/m_e=49$. What happens in a
real plasma with  $m_p/m_e=1836$? 
As shown in Fig.~\ref{growth} and Fig.~\ref{dispersion}, the effect of an
increased proton mass is wave growth at larger $k$. Additionally the
maximum growth rate will be slightly larger. 
 
The shift of the maximum growing wave number for reduced anisotropy, 
as well as the reduction of the wave frequencies at fixed wave numbers,
will also occur in a plasma with real mass ratio. 
E.g. a reduction of the electron temperature ratio from $T^e_\|/T^e_\perp
\approx 20$ to $T^e_\|/T^e_\perp \approx 6$  will shift the wave number of
maximum growth from $kc/\omega_p=1.2$ to $kc/\omega_p=1.8$.  

What happens to the protons? Keeping the ratio $\Omega_e/\omega_e$
constant, the Alfv\'en-speed $c_A$ has to be scaled by $\sqrt{m_r}$,
where $m_r$ is the ratio between artificial and real proton mass. 
At the same time, the thermal velocity has to be scaled by 
$\sqrt{m_r}$, thus the ratio $v_\|/v^p$ remains constant. 
However, for a fixed wave number, the frequency
decreases due to the different dispersion relation. 
It was shown in  Sect.~\ref{section:proton}, that the fastest 
growing wave was in resonance with protons much slower than the 
thermal velocity. This is still true in case of a plasma 
with real mass ratio, e.g. for a temperature ratio of
$T^e_\|/T^e_\perp\approx6$, the fastest growing mode is 
$(kc/\omega_p;\omega/\Omega_p)=(1.7;0.9)$ and 
the resonance velocity is $v_\|/v^p=0.16$. 

Due to the resonant absorption of the waves by the protons and the
broader wave spectrum excited in a real plasma, the heating of the
protons due to the EFI may be larger than predicted by the simulations.
Additionally it was demonstrated in the previous section, that the 
absorption of field energy by the protons has only minor influence 
on the electrons. 
As the only drain of free electron energy is either isotropization 
or wave growth, and a broader spectrum of waves is excited  in a real
plasma,  the simulated electron temperature isotropization time can 
be assumed to be an upper limit. 
Faster isotropization is confirmed by a simulation with 
increased mass ratio of $m_p/m_e=81$ (see Fig.~\ref{instab}).  

\section{Summary}
\label{section:summary}
The preferred acceleration along an external magnetic field compared to
the perpendicular direction causes an unsolved  problem in solar 
flare particle acceleration mechanisms.
Without additional scattering mechanisms, this leads to an anisotropic 
electron velocity distribution, which can 
in turn reduce the efficiency of the accelerator. The isotropization
time of an anisotropic velocity distribution has to be
short compared to the total acceleration time which is
$0.4$ s or less to energize the electrons to $20$ keV.   

Based on linear theory, the Electron Firehose Instability (EFI) has 
been proposed as a candidate for velocity space particle scattering.
In the previous sections, the temporal development of the EFI and 
its influence on the particle velocity distribution has been shown,  
based on self-consistent electromagnetic PIC simulations. 
The plasma parameters were chosen to be  comparable to expected values 
in a  solar flare plasma. 

In case of a rod shaped geometry, which allows long-wavelength
modes to grow mainly parallel to the external magnetic field, the process
responsible for isotropization is identified to be the Electron
Firehose Instability. The identification is made by comparison of  the 
growth rate and the dispersion relation between simulations and linear
theory.

The EFI is driven by the bulk of the electron velocity distribution, 
and is thus non-resonant. Most of the parallel electron energy is thereby
converted into perpendicular electron energy. 
However, some free energy is converted into magnetic field energy.

The growing waves are left hand polarized electromagnetic waves with
frequencies close to $\Omega_p$. 
As soon as the wave amplitudes are large enough, they start to scatter the 
electrons in velocity space by anomalous Doppler resonance. 
In this phase, the electrons emit most of their free energy as 
waves by reducing quickly their anisotropy.
Wave growth stops after the anisotropy has been reduced below the 
instability criterion. 

The strong waves generated by the instability mainly show up 
in the perpendicular magnetic field. Due to their left hand
polarization, they are normal Doppler resonant with the protons.
The resonance velocities of the waves generated by the EFI
are in the bulk of the proton velocity distribution, and can thus be
easily absorbed by the protons. This leads to perpendicular heating of
the protons. 

At the time of the anomalous Doppler resonance of the electrons with
the wave, the electrons build up bumps in the reduced velocity
distribution. They are the source of Langmuir waves, which in turn may
yield observable radio emission. 

Under solar flare conditions, the simulated plasma corresponds to a plasma
with electrons of 2 keV in perpendicular direction and 23keV
parallel to the external field. According to the simulations, this 
distribution isotropizes within about $20\Omega_p^{-1}$,
corresponding  to $2\cdot 10^{-5}{\rm s}$. The resulting isotropic 
energy of the electrons is about $7{\rm keV}$.  
The isotropization  is fast compared to the timescales needed to 
accelerate the whole electron distribution to $25$keV, which is
expected to be on many orders of magnitude larger time scales. 
The electron anisotropy originating from acceleration can therefore not
be maintained.

Due to the large growth rates of the EFI at oblique
angles, occurrence of the oblique firehose instability in a real plasma
is even more likely. However, it is not yet clear how much energy is
carried  by those modes and what their influence on the particle
distribution is. Additionally, due to the artificial mass ratio, 
the isotropization time may be overestimated. 
The given isotropization time due to the EFI can  therefore be
considered as an upper limit for an anisotropy to exist. 
Increasing computing resources will allow to tackle the problem in a 
slap shaped or even fully 3D geometry with larger mass ratios. 
This is the topic of current investigations. 

In these simulations, most of the free energy is used to heat 
the electrons in perpendicular
direction. Little  energy goes into the protons, heating them in
perpendicular direction and cooling them in parallel direction. The
isotropization process can therefore not be used to explain bulk
heating of the protons. 

For the assumed plasma parameters, the Electron Firehose Instability
limits the anisotropy. The particle distribution can therefore be
assumed to be isotropic throughout the acceleration process.

\begin{acknowledgement}
Access to the {\em Asgard} Beowulf-Cluster was granted by the Institute of
Theoretical Physics at ETH Z\"urich. Utilization of the WHAMP code was 
simplified by G. Paesold's user interface IDLWhamp. Special thanks to 
A. O. Benz for many fruitful discussions. Additional thanks to an
unknown referee for many  suggestions to improve the paper.
Parts of this work were financially supported by the Swiss National
Science foundation, Grant No. 20-536664.98.
\end{acknowledgement}

\end{document}